\documentclass[twocolumn,prl,showpacs,amssymb,superscriptaddress,amsmath]{revtex4}

\usepackage{bm}
\usepackage{graphicx}
\usepackage{color}
\usepackage{dcolumn}

\begin{document}

\title{Recoil effects of photoelectrons in a solid}

\author{Y. Takata}
\email{E-mail: takatay@spring8.or.jp}
\affiliation{Soft X-ray Spectroscopy Laboratory, RIKEN SPring-8 Center, Sayo-cho, Hyogo 679-5148, Japan}
\author{Y. Kayanuma}
\email{E-mail: kayanuma@ms.osakafu-u.ac.jp}
\affiliation{Graduate School of Engineering, Osaka Prefecture University, Sakai, Osaka, 599-8531 Japan}
\author{M. Yabashi}
\affiliation{Coherent X-ray Optics Laboratory, RIKEN SPring-8 Center, Sayo-cho, Hyogo 679-5148, Japan}
\affiliation{JASRI, SPring-8, Sayo-cho, Hyogo 679-5198, Japan}
\author{K. Tamasaku}
\affiliation{Coherent X-ray Optics Laboratory, RIKEN SPring-8 Center, Sayo-cho, Hyogo 679-5148, Japan}
\author{Y. Nishino}
\affiliation{Coherent X-ray Optics Laboratory, RIKEN SPring-8 Center, Sayo-cho, Hyogo 679-5148, Japan}
\author{D. Miwa}
\affiliation{Coherent X-ray Optics Laboratory, RIKEN SPring-8 Center, Sayo-cho, Hyogo 679-5148, Japan}
\author{Y. Harada}
\affiliation{Soft X-ray Spectroscopy Laboratory, RIKEN SPring-8 Center, Sayo-cho, Hyogo 679-5148, Japan}
\author{K. Horiba}
\affiliation{Soft X-ray Spectroscopy Laboratory, RIKEN SPring-8 Center, Sayo-cho, Hyogo 679-5148, Japan}
\author{S. Shin}
\affiliation{Soft X-ray Spectroscopy Laboratory, RIKEN SPring-8 Center, Sayo-cho, Hyogo 679-5148, Japan}
\author{S. Tanaka}
\affiliation{Graduate School of Sciences, Osaka Prefecture University, Sakai, Osaka, 599-8531, Japan}
\author{E. Ikenaga}
\affiliation{JASRI, SPring-8, Sayo-cho, Hyogo 679-5198, Japan}
\author{K. Kobayashi}
\affiliation{JASRI, SPring-8, Sayo-cho, Hyogo 679-5198, Japan}
\author{Y. Senba}
\affiliation{JASRI, SPring-8, Sayo-cho, Hyogo 679-5198, Japan}
\author{H. Ohashi}
\affiliation{JASRI, SPring-8, Sayo-cho, Hyogo 679-5198, Japan}
\author{T. Ishikawa}
\affiliation{Coherent X-ray Optics Laboratory, RIKEN SPring-8 Center, Sayo-cho, Hyogo 679-5148, Japan}
\affiliation{JASRI, SPring-8, Sayo-cho, Hyogo 679-5198, Japan}

\date{\today}

\begin{abstract}
\textbf{ABSTRACT}

High energy resolution C 1$s$ photoelectron spectra of graphite were measured at the excitation energy of 340, 870, 
5950 and 7940eV using synchrotron radiation. On increasing the excitation energy, i.e., increasing kinetic 
energy of the photoelectron, the bulk origin C 1$s$ peak position shifts to higher binding energies. This systematic shift is due to 
the kinetic energy loss of the high-energy photoelectron by kicking the atom, and is clear evidence of the recoil 
effect in photoelectron emission. It is also observed that the asymmetric broadening increases for the higher 
energy photoelectrons. All these recoil effects can be quantified in the same manner as the M\"ossbauer effect for 
$\gamma$-ray emission from nuclei embedded in crystals.\end{abstract}
\maketitle

%############## Introduction ########################

Photoelectron spectroscopy is widely used for the study of electronic structure of solids~\cite{Huffner}. 
The binding energy $E_{B}$ of the electron is often calculated from the photoelectron kinetic energy using the following equation:
\begin{equation*}
E_{B} = h\nu - E_{kin} - \phi,
\end{equation*}
where $E_{kin}$ is the measured electron kinetic energy, $h\nu$ is the photon energy for excitation, 
and $\phi$ is the work function. This procedure overlooked 
recoil effects, which lead to part of the kinetic energy being imparted to the emitting atom. As a result, the binding energy determined in this way 
is greater than the true binding energy. This effect is very small for vacuum ultraviolet and soft x-ray photoelectron spectra, so that recoil 
effects can be safely neglected. For photoelectrons with 1000 eV kinetic 
energy emitted from carbon atom, the recoil energy is estimated to be only $\sim$ 45 meV, although Kukk \textit{et al.} succeeded recently to observe 
small deviation in spectral shape of vibronic lines in gaseous methane, which have been attributed to a recoil effect~\cite{Kukk}.
\par
The momentum transfer at recoil is a fundamental process observed in experiments of neutron and x-ray scattering 
~\cite{Kittel}, high-energy electron backscattering~\cite{Laser,Werner}, etc. For photoelectron emission, Domcke and Cederbaum~\cite{Domcke} 
predicted that the 
recoil effect can be observed as a spectral modification for gaseous molecules with light atoms. 
Quite recently, Fujikawa {\textit{et al.}~\cite{Fujikawa} evaluated the amount of shift and broadening of core-level 
photoelectron spectra, as well as for electron backscattering, due to recoil effect in solids. 
It is noted that at keV energies, since the momentum of an electron is much larger than that 
of a photon of the same energy, and the transferred momentum is largely that of the emitted electron, 
it should be possible to detect recoil effects in photoelectron emission with keV energies.

In the last few years, hard x-ray photoelectron spectroscopy with the excitation energy of 
6-8 keV has been realized using high brilliance synchrotron radiation~\cite{Kobayashi, Takata01, VOLPE},
resulting in useful studies on semiconductors and correlated materials~\cite{Takata02, Taguchi}. 
Since the achieved energy resolution is quite good ($\Delta E<80$ meV), it gives us an opportunity 
to investigate recoil effects in a solid.

In this study, we measured the C 1s core 
level photoelectron spectra of highly oriented pyrolytic graphite (HOPG), 
using high energy resolution ($\Delta E=100 \sim 120$ meV), with the aim of investigating 
recoil effects due to photoelectron emission in a solid. HOPG is a suitable material for study of the 
recoil effect because of the light atomic mass of carbon. Furthermore, reliable studies of 
surface-bulk core level shift of HOPG have already 
been established recently~\cite{Sette,Prince, Balas, Smith}, providing a very suitable reference for this study.
The experimentally obtained spectra in the present study confirm the surface core level shift.
For the bulk derived feature, the spectra exhibit systematic kinetic energy loss and 
the anomalous asymmetric broadening on increasing the excitation energy. All these features are clear evidence of recoil effect, and  
can be theoretically quantified in the same picture as the M\"ossbauer effect for $\gamma$-ray emission.

%###########  Experimental #############################
Measurements of C 1$s$ photoelectron spectra of HOPG were performed at SPring-8 using 
synchrotron radiation. Hard x-ray photoelectron spectra at the excitation energy of 5950 and 
7940 eV were measured at the undulator beamline BL29XU using a hemispherical electron energy 
analyzer, SCIENTA R4000-10kV. Details of the apparatus including x-ray optics are 
described in refs.~\cite{Takata01, Tamasaku, Ishikawa}. Soft x-ray spectra at 
the excitation energy of 340, 870 eV were measured at the undulator beamline BL17SU~\cite{Ohashi} using 
an electron analyzer, SCIENTA SES-2002. 
Clean surfaces of HOPG were prepared by peeling off an adhesive tape at a pressure of 
~10$^{-8}$ Pa for all measurements. All the measurements on graphite reported here
were carried out at room temperature. The energy scale of the spectra were calibrated 
very accurately ($<$10 meV) by measurements of the Au 4f core levels and the Fermi edge
at room temperature. The total instrumental energy resolution ($\Delta E$) for 
the soft x-ray and hard x-ray spectra were determined as 100 meV and 120 meV by fitting the Fermi-edge 
profiles of Au measured at 20 K. 

%###########  Results and discussion #############################
C 1$s$ core level spectra of HOPG measured at room temperature with soft x-ray excitation 
(340 and 870 eV) and hard x-ray excitation (5950 and 7940eV) are shown in Fig. 1(a).
The spectra are obtained effectively in normal emission geometry. 
The photoelectron detection angle relative to the sample surface for the 
soft x-ray and hard x-ray spectra were 90\r{ } and 85\r{ }, respectively.  
The peak position of the spectrum obtained with $h\nu$=870 eV shifts to lower binding energy
(284.39 eV)in comparison with that of $h\nu$=340 eV spectrum (284.47 eV). This shift is due to the difference 
in the probing depth of photoelectrons between these excitation energies. 
With increase of the kinetic energy, the probing depth of a photoelectron 
becomes larger. The spectrum of $h\nu$=340 and $h\nu$=870 eV are dominated by the surface 
and bulk components, respectively. The observed surface-bulk core level 
shift is consistent with the results reported by Balasubramanian et al.~\cite{Balas},
using photon energies of 300-348 eV, and a total energy resolution of about 50 meV.
It was shown that surface-bulk core level splitting results in a weak bulk feature 
at a lower binding energy compared to the surface derived feature. 
In the present study, our $h\nu$=340 eV spectrum is similar to that reported by Balasubramanian et al.~\cite{Balas}, 
while the $h\nu$=870 eV spectrum is dominated by the bulk derived feature.

\begin{figure}
\centering
\includegraphics[height=10cm]{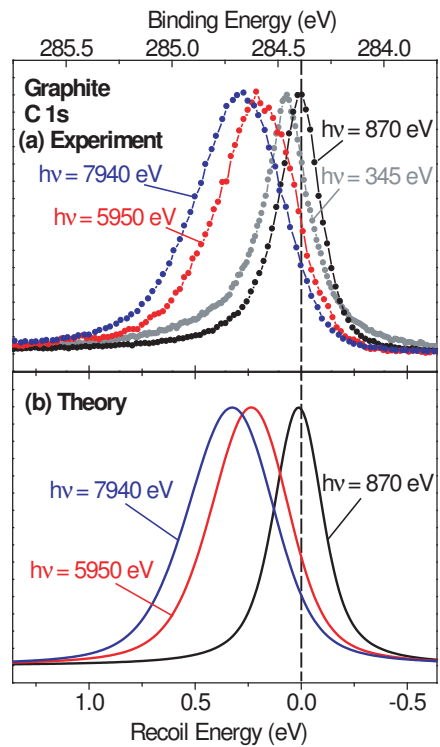}
\caption{(color online)(a) Photon energy dependence of C 1$s$ core level spectra of graphite. 
The soft x-ray ($h\nu=340, 870$ eV) and hard x-ray ($h\nu=5950, 7940$ eV) are measured at 
the emission angle of 90\r{ } and 85\r{ } relative to the sample surface. 
(b) Theoretically obtained spectra taking into account the recoil effect 
in a Debye model with $\hbar\omega_{b,D}=$75 meV.}

\label{tfig:1}       % Give a unique label
\end{figure}

With increase of the excitation energy, i.e., the kinetic energy of the C 1$s$ photoelectron, 
the peak position shifts to higher binding energy side. It is also obvious that 
asymmetric broadening becomes much wider for the higher energy photoelectrons, while 
the total instrumental energy resolution for these spectra is almost the same. 
For the possible origin of these spectral changes, the effect of associated elementary 
excitation such as plasmons is discarded, since the peak position itself shifts, 
depending on the kinetic energy. The asymmetric line shape of the C 1s core level spectra obtained with soft x-ray excitation 
has been discussed in relation with semimetallic character of graphite, and can be fitted 
by the Doniach-Sunjic function~\cite{Sette,Prince, Balas, Smith}. This possibility can be also 
excluded for the same reason. Thus, we are led to a picture based on recoil 
effects for explaining the observed peak shift
and broadening, which depends on the kinetic energy of the photoelectron.

For an atom with mass $M$ in free space, the recoil energy $\delta E$ is simply estimated from the momentum conservation 
as $\delta E=(m/M)E_{kin}$, where $m$ is 
the electron mass. For a carbon atom with the mass ratio $m/M=1/22000$, $\delta E$ 
becomes as large as 0.36eV for $E_{kin}=$8keV. In the solid, this recoil energy is absorbed 
by the phonon bath, resulting in the excitation of phonons. The zero-phonon transition 
corresponds to the event in which the recoil energy is transferred to the center of mass 
motion of the total crystal. This is essentially the same as the M\"ossbauer effect 
in the  $\gamma$-ray emission from nuclei embedded in crystals~\cite{Mossbauer}.

Without loss of generality, we assume that 
a core electron of the carbon atom located at the lattice point $\vec{R}^0$ is ejected by 
the x-ray irradiation. The actual nuclear position $\vec{R}$ may deviate from its equilibrium 
value $\vec{R}^0$ as $\vec{R}=\vec{R}^0+\vec{u}$ because of the thermal and zero point vibrations. 
The interaction Hamiltonian with the x-ray is given by
%\begin{equation*}
$
H_I = (a+a^\dagger) \vec{\varepsilon } \cdot \vec{p},
$
%\end{equation*}
aside from irrelevant factors, where $a$ is the annihilation operator for the x-ray 
with energy $h\nu$ and the polarization vector $\vec{\varepsilon }$. The momentum of photon is neglected here, 
since it is an order of magnitude smaller than the that of the emitted electron in this x-ray energy region. The initial state of transition is given by
%\begin{equation}
$
|\Psi_i \rangle = |h\nu \rangle \otimes |\psi_c \rangle \otimes |i \rangle ,
$
%\end{equation}
where $|h\nu \rangle$  is the one-photon state, $|\psi_c \rangle$ is the core electron 
state with energy $\varepsilon_c$, and $|i \rangle$  is a phonon state of the crystal. The crucial point of the theory is that the wave function 
of the core electron is given by the form
$
\langle \vec{r} \;| \psi_c \rangle = \psi_c (\vec{r}-\vec{R}).
$
This is the adiabatic approximation, and the recoil effect results directly from this 
functional form. On the other hand, the final state of the transition is given by
%\begin{equation}
$
|\Psi_f \rangle = |0\rangle \otimes |\psi_k \rangle \otimes |f\rangle ,
$
%\end{equation}
in which $|0\rangle$ is the vacuum of the photon, $|\psi_k \rangle$ is the plane wave of the electron,
$
\langle \vec{r}\;| \psi_k \rangle = (2\pi)^{-3/2}\exp ( i\vec{k} \cdot \vec{r}),
$
with energy $\hbar^2\vec{k}^2/2m$ , and $|f\rangle$ is a phonon state.

Using the functional forms for $|\Psi_i \rangle$ and $|\Psi_f \rangle$, and changing the integration variable from $\vec{r}$ to $\vec{r}-\vec{R}$ , 
we find
\begin{equation*}
\left\langle \Psi_f |H_I|\Psi_i \right\rangle = \vec{\varepsilon}\cdot\vec{\mu} \left\langle 
f \left|e^{-i\vec{k}\cdot \vec{R}}\right|i \right\rangle, 
\end{equation*}
in which
\begin{equation*}
\vec{\mu} = (2\pi)^{-3/2}\int\!d^3 re^{-i\vec{k}\cdot\vec{r}} \left( -i\hbar \frac{\partial}{\partial \vec{r}} \right) \psi_c (r).
\end{equation*}
The above expression of the interaction Hamiltonian coincides with that for the M\"ossbauer effect~\cite{Maradudin}, 
so that we readily obtain the expression for the photoelectron spectrum as a function of the relative binding energy $E$ measured from the 
recoilless value,
\begin{equation*}
I(E) = \frac{|\vec{\varepsilon}\cdot\vec{\mu}|^2}{2\pi} \int_{-\infty}^{\infty}dt e^{-iEt/\hbar -\Gamma |t|/\hbar}F(t),
\end{equation*}
in which the generating function $F(t)$ is given by the canonical average,
\begin{equation*}
F(t) = \left \langle e^{i\vec{k}\cdot\vec{u}(t)} e^{-i\vec{k}\cdot\vec{u}} \right \rangle ,
\end{equation*}
with $\vec{u}(t)$ being the Heisenberg representation of $\vec{u}$  at time $t$, and $\Gamma$ is the lifetime broadening factor.
For harmonic crystals, $F(t)$ can be written in the closed form, $F(t)=\exp[G(t)]$ where
\begin{equation*}
G(t) = \sum_q \alpha_q^2 \left\{ 
        \left(
                2n(\omega_q)+1
        \right)
        \left(
                \cos \omega_q t -1
        \right)
-i \sin \omega_q t
\right\},
\end{equation*}
with
\begin{equation*}
\alpha_q^2 = \left(
        \frac{\hbar}{2N\tilde{M}\omega_q} 
\right)\left|
                \vec{k} \cdot \vec{\eta_q}
        \right|^2,
\end{equation*}
and
\begin{equation*}
n(\omega_q) = 1/\left( e^{\hbar \omega_q/k_B T} -1 \right),
\end{equation*}
in which $q$ is the abbreviation for the wave vector and the branch index of phonons, $\tilde M$ is the mass of unit cell, $N$ is the number of 
unit cells, and $\vec \eta_q$ is the polarization vector of the phonon.
The photoelectron spectrum $I(E)$ is the convolution of a structure function 
with the Lorentz function. The former consists of the recoilless line (the zero-phonon line) 
and its phonon side bands. Unlike the M\"ossbauer effect, one can control the so-called 
Debye-Waller factor continuously from almost recoilless to strong coupling by changing the 
energy of the x-ray. The spectrum may depend also on the relative angle of 
the $\vec{k}$ vector of the emitted electron because of the anisotropy of the phonon spectrum. 

In actual calculation, we adopt an anisotropic Debye model for graphite. The stretching (in-plane) 
mode and the bending (out-of-plane) mode are assumed to be independent. The surfaces of constant frequency for each mode 
have an prolate spheroidal form in $\vec q$ space reflecting the highly anisotropic dispersion relation for the in-plane propagation 
and the out-of-plane propagation\cite{Krumhansl}. Then, $G(t)$ is given by 
\begin{equation}
G(t)=\int_{-\infty}^\infty d\omega \bigr(e^{i\omega t}-1\bigr)\bigl\{J_s(\omega)\cos^2\theta+J_b(\omega)\sin^2\theta\bigr\}
\end{equation}
for the emission angle $\theta$ relative to the surface, where $J_\lambda(\omega)\quad (\lambda=s,b)$ are given by
\begin{equation*}
J_\lambda(\omega)=\frac{\delta E}{\hbar\omega}\bigl\{\bigl(n(\omega)+1)D_\lambda (\omega)\Theta(\omega)
+n(|\omega|)D_\lambda(|\omega|)\Theta(-\omega)\bigr\}
\end{equation*}
with the step function $\Theta(\omega)$. Here, $D_\lambda (\omega)$ is the density of state per atom, 
\begin{eqnarray}
D_\lambda(\omega)=\frac{6}{3\omega_{\lambda,D}^2-\omega_{\lambda,C}^2}\left\{
\begin{array}{ll}
\frac{\omega^2}{\omega_{\lambda,C}}, &0 < \omega < \omega_{\lambda,C}\\
\omega, & \omega_{\lambda,C} < \omega < \omega_{\lambda, D}\\
0, & {\rm otherwise}.
\end{array}
\right.
\end{eqnarray}

The density of state has a characteristic of three-dimensional Debye model in low frequency $(\omega < \omega_{\lambda, C})$, 
but of two-dimensional one at high frequency, where only the in-plane propagating modes contributes to $D_{\lambda}(\omega)$. 
The Debye cut-off frequencies for the stretching mode $\hbar\omega_{s, D}$ and the bending mode $\hbar\omega_{b, D}$ are  
estimated from the frequency at $K$-point of LA mode and ZA mode, respectively. From the 
experimental dispersion curves\cite{Wirtz}, we fix, $\hbar\omega_{s, D}=$150meV, $\hbar\omega_{b, D}=$75meV, 
$\hbar\omega_{s, C}=$6.3meV, $\hbar\omega_{b, C}=$12.5meV. 

\begin{figure}
\centering
\includegraphics[height=10cm]{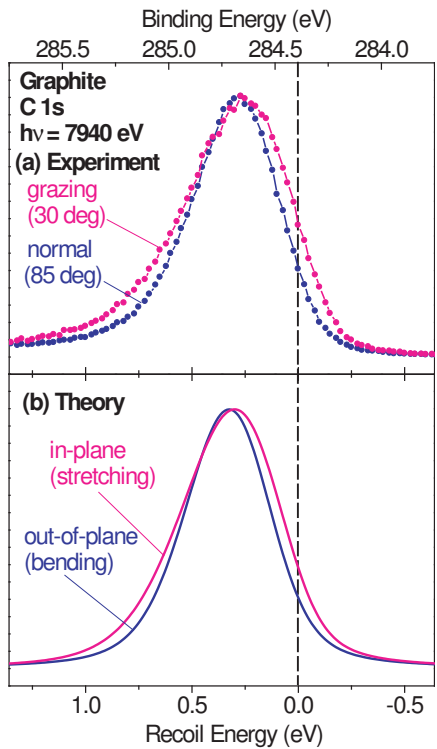}
\caption{(color online)(a) C 1$s$ core level spectra of graphite measured at the photon energy of 
7940 eV at the emission angle of 85\r{ } (normal) and 30\r{ } (grazing) relative to the 
sample surface. (b) Theoretically obtained spectra with the Debye energy $\hbar\omega_{b,D}$=75 meV and $\hbar\omega_{s,D}$=150 meV 
for the bending and the stretching mode, respectively.}

\label{tfig:2}       % Give a unique label
\end{figure}
The theoretical spectra of the C 1$s$ normal emission photoelectrons 
of graphite were calculated and are plotted as a function of the recoil energy in Fig. 1(b). 
The lifetime broadening (full-width at half maximum) is taken to be 160 meV as is known for graphite~\cite{Prince, Balas}. 
The spectra are then convoluted with a Gaussian function corresponding to the experimental resolution of 120 meV.    
Theoretical spectra reproduce the experimental spectra 
fairly well without any adjustable parameters. Note that, at 870 eV excitation, the  peak shift is quite 
small, but phonon excitations cause the asymmetric broadening.
 
Fig. 2 (a) shows the experimental C 1s core level spectra of HOPG measured with 7940 eV 
excitation at the emission angle of 85\r{ } (the same as that in Fig. 1(a)) and 30\r{ } relative to 
the sample surface. The peak slightly shifts to lower binding energy and becomes broader for the 
grazing emission spectrum. Theoretical spectra are shown in Fig. 2 (b). The theoretical spectra 
reproduce well the observed emission angle dependence. The larger spectral width in the grazing angle 
emission than the normal one is attributed essentially to the large Debye cut-off frequency $\omega_{s, D}$ for the stretching mode, 
which is roughly twice that for the bending mode $\omega_{b, D}$.

It is remarkable that the gross spectral features in Fig. 1 can be understood by a semiclassical approximation~\cite{Fujikawa}. 
We have checked that the spectra can be approximately fitted by a Gaussian line shape, using an atomic value for the 
shift $\delta E$, and a width given by the second moment due to the Doppler broadening of $2\delta E k_BT$.
However, the semiclassical picture fails in 
reproducing the asymmetries of the observed line shapes, which are signatures of the 
quantum nature of phonons. The anisotropic line shape shown in Fig. 2 also indicates the necessity of a full quantum 
mechanical analysis taking into account the solid state effect for such a high resolution spectroscopy. 
Furthermore, the good agreement between the experimental data and the theoretical calculation tells us that 
the multiple-scattering effect is small.

In conclusion, recoil effects in photoelectron emission from solid were observed in the high energy 
C 1$s$ core level spectra of graphite. The observed spectral shapes are successfully quantified in the 
same manner as the M\"ossbauer effect for $\gamma$-ray emission from nuclei embedded in crystals. 
The recoil effect exists, and is measurable, making it an important aspect of high-energy 
photoelectron spectroscopy.

%Acknowledgement
We acknowledge Dr. Chainani for his valuable comments and critical reading of the manuscript.
This work was partially supported by the Ministry of Education, Science, Sports and Culture 
through a Grant-in-Aid for Scientific Research (No. 15206006 and 18540323).  

%###################################################
%\newpage
%\begin{center}
%\large\textbf{REFERENCES}
%\end{center}

%\clearpage

\begin{thebibliography}{}
    
\bibitem{Huffner}
See, for example, S. H\"ufner: \textit{Photoelectron Spectroscopy}, 3rd edn (Springer-Verlag, Berlin-Hidelberg 2003)

\bibitem{Kukk}
E. Kukk \textit{et al.}, Phys. Rev. Lett. {\bf 95}, 133001 (2005).

\bibitem{Kittel}
See, for example, C. Kittel: \textit{Quantum Theory of Solids}. (John Wiley and Sons, New York 1963), Chap. 19.

\bibitem{Laser}
D. Laser and M. Seah, Phys. Rev. B{\bf 47}, 9836 (1993).

\bibitem{Werner}
W. S. Werner, C. Tomastic, T. Cabela, G. Richter, and H. St\"ori, J. Elect. Spect. Related Phenom. {\bf 113}, 127 (2001).

\bibitem{Domcke}
W. Domcke and L. S. Cederbaum, J. Elect. Spect. Related Phenom. {\bf 13}, 161 (1978).

\bibitem{Fujikawa}
T. Fujikawa, R. Suzuki, and L. K\"over, J. Elect. Spect. Related Phenom. {\bf 151}, 170 (2006).
    
\bibitem{Kobayashi} 
K. Kobayashi \textit{et al.}, Appl. Phys. Lett. {\bf 83}, 1005 (2003).

\bibitem{Takata01} 
Y. Takata \textit{et al.}, Nucl. Instrum. Methods. A {\bf 547}, 50 (2005).

\bibitem{VOLPE} 
P. Torelli \textit{et al.}, Rev. Sci. Instrum. {\bf 76}, 23909 (2005).

\bibitem{Takata02} 
Y. Takata \textit{et al.}, Appl. Phys. Lett. {\bf 84}, 4310 (2004).

\bibitem{Taguchi}
M. Taguchi \textit{et al.}, Phys. Rev. Lett. {\bf 95}, 177002 (2005).

\bibitem{Sette}
F. Sette, \textit{et al.}, Phys. Rev. B {\bf 41}, 9766 (1990).

\bibitem{Prince}
K. C. Prince, \textit{et al.}, Phys. Rev. B {\bf 62}, 6866 (2000).


\bibitem{Balas}
T. Balasubramanian, J. N. Andersen and L. Walld\'en, Phys. Rev. B {\bf 64}, 205420 (2001).

\bibitem{Smith}
R.A. P. Smith \textit{et al.}, Phys. Rev. B {\bf 66}, 245409 (2002).

\bibitem{Tamasaku}
K. Tamasaku \textit{et al.}, Nucl. Instrum. Methods. A {\bf 467/468}, 686 (2001).

\bibitem{Ishikawa}
T. Ishikawa, K. Tamasaku and M. Yabashi, Nucl. Instrum. Methods. A {\bf 547}, 42 (2005).

\bibitem{Ohashi}
H. Ohashi \textit{et al.}, to be published in AIP Conference Proceedings on 9th International 
Conference on Synchrotron Radiation Instrumentation (SRI 2006).

\bibitem{Mossbauer}
R. L. M\"ossbauer, Z. Physik {\bf 151}, 124 (1958).

\bibitem{Maradudin}
A. A. Maradudin, P. A. Flinn,and J. M. Radcliffe, Annals Phys. {\bf 26}, 81 (1964).

\bibitem{Krumhansl}
J. Krumhansl and H. Brooks, J. Chem. Phys. {\bf 21}, 1663 (1953).

\bibitem{Wirtz}
L. Wirtz and A. Rubio, Solid State Commun. {\bf 131}, 141 (2004), and references cited therein.



\end{thebibliography}
\end{document}